# Plasma engineering of microstructured piezo – triboelectric hybrid nanogenerators for wide bandwidth vibration energy harvesting

Xabier García-Casas,[1] Ali Ghaffarinehad,[1]* Francisco J. Aparicio,[1]* Javier Castillo-Seoane,[1,2] Carmen López-Santos,[1,3] Juan P. Espinós,[1] José Cotrino,[1,2] Juan Ramón Sánchez-Valencia,[1,2] Ángel Barranco[1] and Ana Borrás[1]*

1 Nanotechnology on Surfaces and Plasma Group (CSIC-US), Materials Science Institute of Seville (Consejo Superior de Investigaciones Científicas – Universidad de Sevilla) c/Américo Vespucio 49, 41092, Seville, Spain.

2 Departamento de Física Atómica, Molecular y Nuclear, Universidad de Sevilla, Avda. Reina Mercedes, E-41012, Seville, Spain.

3 Department of Applied Physics I, University of Seville, c/ Virgen de Africa 7, 41011, Seville, Spain.

Anaisabel.borras@icmse.csic.es; ali.ghaffarinejad@icmse.csic.es; fjaparicio@icmse.csic.es

**ABSTRACT:** We introduce herein the advanced application of low-pressure plasma procedures for the development of piezo and triboelectric mode I hybrid nanogenerators. Thus, plasma assisted deposition and functionalization methods are presented as key enabling technologies for the nanoscale design of ZnO polycrystalline shells, the formation of conducting metallic cores in core@shell nanowires, and for the solventless surface modification of polymeric coatings and matrixes. We show how the perfluorinated chains grafting of polydimethylsiloxane (PDMS) provides a reliable approach to increase the hydrophobicity and surface charges at the same time that keeping the PDMS mechanical properties. In this way, we produce efficient Ag/ZnO convoluted piezoelectric nanogenerators supported on flexible substrates and embedded in PDMS compatible with a contact–separation triboelectric architecture. Factors like crystalline texture, ZnO thickness, nanowires aspect ratio, and surface chemical modification of the PDMS are explored to optimize the power output of the nanogenerators aimed for harvesting from low-frequency vibrations. Just by manual triggering, the hybrid device can charge a microcapacitor to switch on an array of color LEDs. Outstandingly, this simple three-layer architecture allows for harvesting vibration energy in a wide bandwidth, thus, we show the performance characteristics for frequencies between 1 Hz to 50 Hz and demonstrate the successful activation of the system up to ca. 800 Hz.

Piezoelectric and triboelectric hybrid nanogenerators (PTNGs) have been reported during the last years as an opportunity to harvest kinetic energy from vibrations and mechanical deformations. These hybrid nanogenerators offer the high voltage outputs characteristic from triboelectric nanogenerators (TENGs) along with the quick response to deformations and higher current densities from piezoelectric nanogenerators (PENGs).[1-5] Besides, hybrid mechanical (vibration-base) harvesters including diverse combinations of electrostatic, magnetostrictive, tribo, piezo, and electromagnetic nanogenerators have been proposed as a solution to overcome a major challenge of energy scavenging, i.e. the conversion from low-frequency (and small deformation) sources which can vary in amplitude and frequency.[2-4] Such is the case for ambient energy sources as body movements, machinery, motor vehicles, moving liquids, wind, etc. Different TENGs and hybrid systems have been reported recently to harvest from wide bandwidth vibrations with frequencies ranging in tens of hertz.[6-15] However, often, the device assembly and operation of these hybrid nanogenerators are complex and not easily extendable from one scenario to the other. According to Z. L. Wang,[1] there are three modes of PENG/TENG coupling, mode I in which PENG and TENG share a pair of electrodes, mode II in which PENG and TENG share one common electrode, and mode III in which PENG and TENG work independently. Herein, we will show early results on the development of a mode I PTHNG for wide frequency harvesting with a simple architecture that consists in the embedding of the piezoelectric material within a triboelectric polymeric matrix, reducing to the minimum the number of layers (three) in the architecture. For this particular application, we have assembled the nanogenerators in vertical contact separation mode but the configuration is highly flexible and fully compatible with other triboelectric architectures as lateral sliding.[16-17] Thus, we show proof of concept devices with ZnO and polydimethylsiloxane (PDMS) as correspondently piezo and triboelectric materials using ITO / PET as substrates and the top electrode. Although the same PTHNG can be adapted to alternative combinations, we have selected ZnO and PDMS as they are widely applied in the fabrication of PENGs and TENGs[1-5,16-20] and offer immediate advantages from both the synthetic and performance points of view.

On one hand, PDMS has been recently demonstrated as the material of choice for negative triboelectric layer because of its advantages as stretchability, optical transparency and biocompatibility, chemical stability, and flexible synthesis by different routes as spin and dip-coating, blading approaches, and bar-assisted printing.[1,16,17,20] The latest results in the literature have revealed several paths for the optimization of PDMS TENGs, as by increasing inner porosity, producing hierarchical microstructure by embossing with templates, laser patterning, or through the implementation of fillers.[20-28] This latter route intends to enhance the electrical performance of the TENGs regarding current output and charges generation. There is also an imperious need to improve the stability of the TENGs working under ambient and harsh conditions, as the presence of moisture and humidity is detrimental for the generation of charges on the negative triboelectric layers.[16,17,20,29] Thus, it is highly desirable to develop hydrophobic or superhydrophobic surfaces that prevent the formation of water or -OH interfacial layer on the polymer. In this regard, the plasma treatment of polymers provides control in their surface tension, roughness, and porosity affecting mechanical properties like adhesion and resistance to erosion and abrasion, and their wettability.[30-33] The advantages of the application of plasma technologies have already impacted the topic of nanogenerators with appealing results in the plasma and Reactive Ion Etching (RIE) treatments of polymeric triboelectric layers,[34-38] fabrication of Diamond Like Coatings (DLCs) and $CF_x$ for TENGs,[39-41] and the recent implications of the combinations of atmospheric plasmas and microplasmas with triboelectric nanogenerators.[42-46] Herein, we will contribute with a phenomenological demonstration on how the oxygen plasma mediated perfluorinated grafting of the PDMS not only increases the water contact angle but also multiplies the power output of the hybrid piezo-triboelectric nanogenerators.

On the other hand, we will also use plasma assisted deposition techniques for the fabrication of the piezoelectric ZnO counterpart.[47,48] Indeed, we will pay special attention to the role played by the nanoscale arrangement of the materials implemented in the devices. The application of ZnO thin films and layers has been recently exploited as an alternative to the single-crystalline counterparts.[18,49,50] In a previous article,[49] we demonstrated the fabrication of polycrystalline ZnO shells and layers by plasma-enhanced chemical vapor deposition (PECVD) as a piezoelectric component in core@shell and multishell nanowires. This was a quite valuable example of how the use of plasma for the fabrication of nanomaterials can be exploited to the competitive and mass-scale development of nanogenerators, in sense of large-area deposition with high homogeneity, low required temperature for the formation of crystalline layers with *adhoc* composition and thickness, high conformality and compatibility with delicate substrates as well as low power consumption and environmental costs relaying on solventless procedures.

The aim of this article is therefore twofold, first, we show how different plasma assisted deposition and processing methods can be applied in the optimization of piezoelectric and triboelectric materials, including the DC sputtering of metal electrodes, the deposition of highly texturized ZnO shells, and the oxygen-plasma mediated grafting of fluorinated molecules. Secondly, we will demonstrate the straightforward fabrication of the piezo and triboelectric hybrid nanogenerator. Indeed, our final objective is to reveal the synergistic combination of the two mechanisms driven by this approach allowing for the development of wide bandwidth vibrations and mechanical strain harvesters. Thus, we will show the response of the system to the excitation at frequencies ranging from a few Hz up to 1000 Hz and relate it to the nature of the hybrid energy harvesting system.

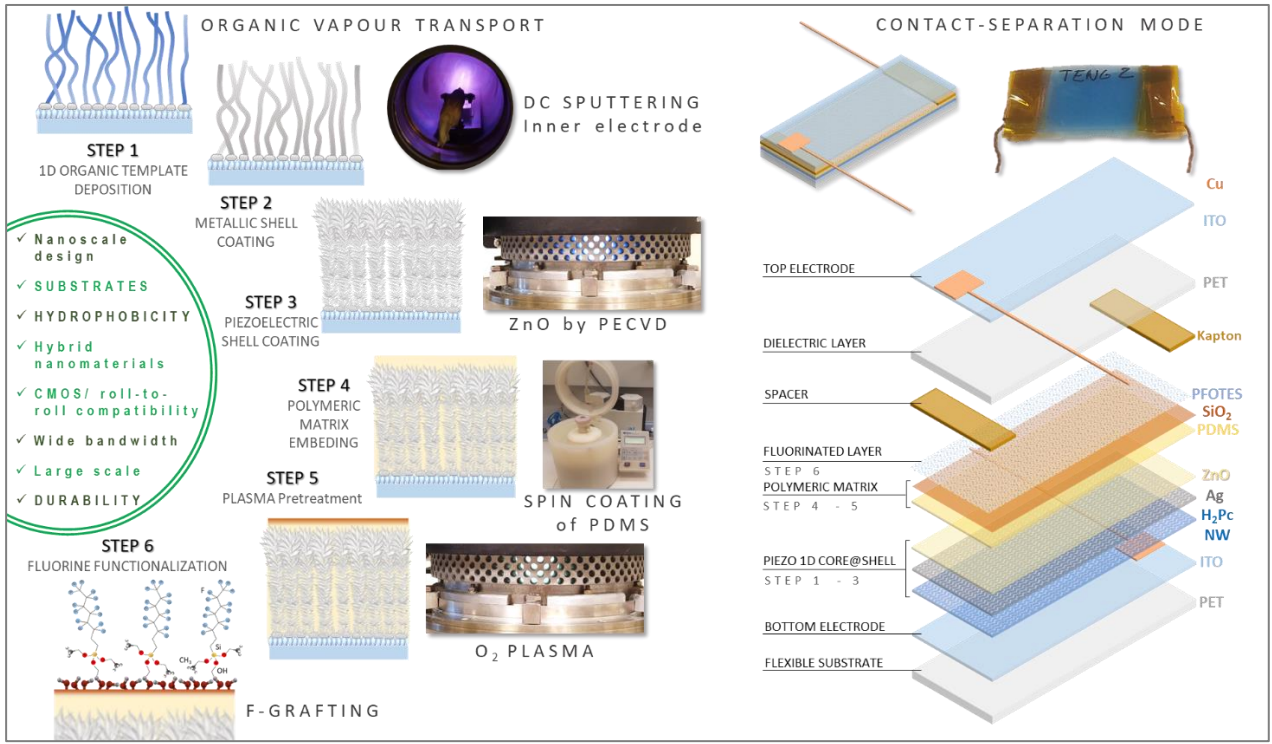

**Schematic 1. Schematics on the device fabrication and assembly.** Left) Step-by-step synthesis of the hybrid core@multishell nanowires and embedding in the PDMS matrix. Right) Layer schematization of the piezoelectric and triboelectric hybrid nanogenerators assembly in vertical separation contact mode, the photograph shows a device of 2 x 4 cm.

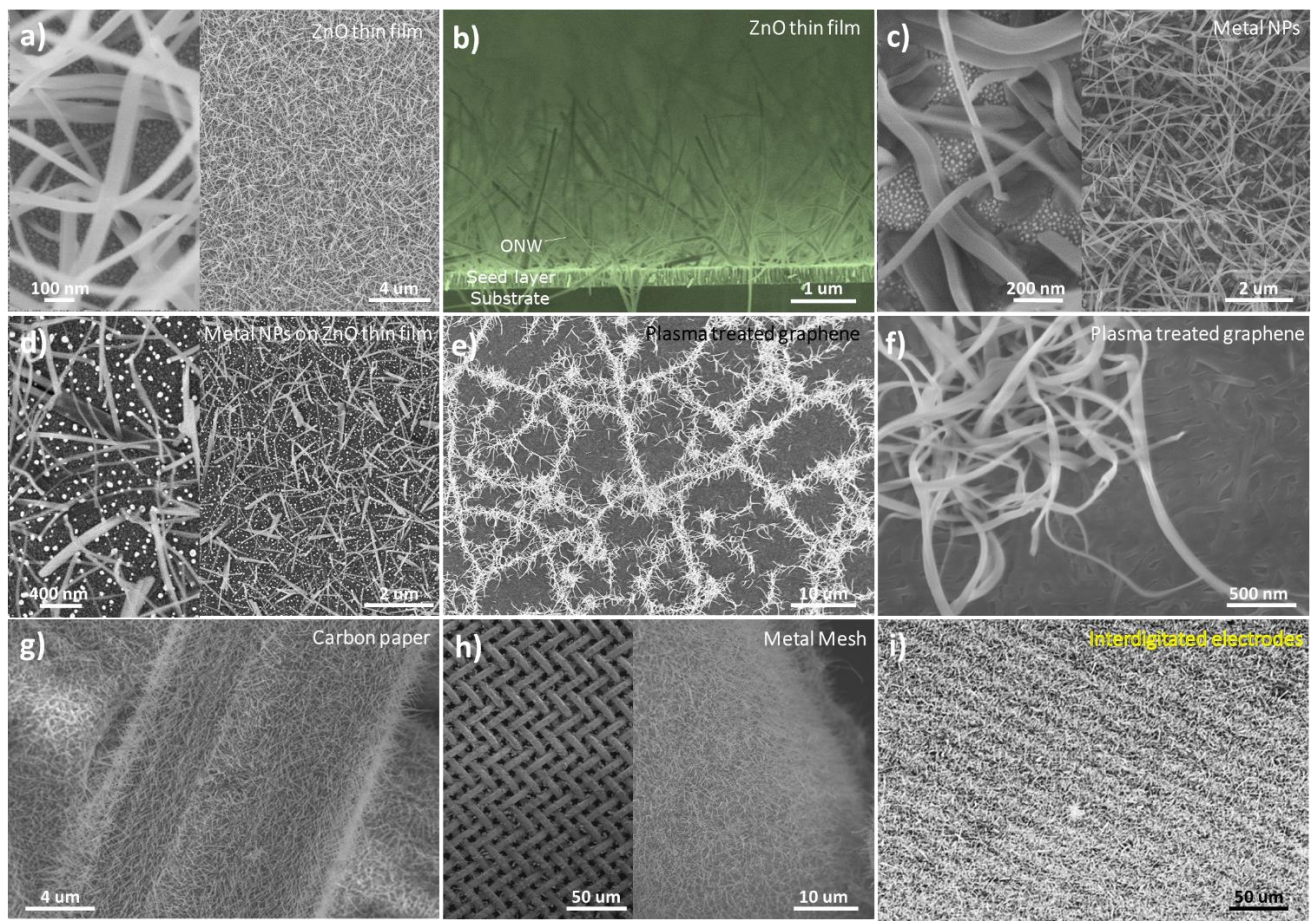

**Figure 1. Universality on the soft-template method, ONWs formed on different substrates.** Characteristic SEM micrographs of the growth of $H_2$-Phthalocyanine nanowires on different substrates, including metal oxide layers (a-b), metal nanoparticles (c) and metal nanoparticles grown on metal oxide thin films (d), $O_2$ plasma treated commercial graphene (e-f), Toray paper (g), stainless steel mesh (h) and interdigitated ITO microelectrodes (i).

**Results and Discussion**

***Synthesis, processing, and physicochemical characterization of the materials in the piezo/tribo hybrid nanogenerators.***

Schematic 1 gathers the different steps and device architectures developed to increase the power output delivered by the hybrid nanogenerators consisted of piezoelectric core@multishell nanowires embedded in surface-treated PDMS working in vertical contact separation mode architecture. The left part of the schematic shows the synthetic multistep procedure developed for the fabrication of core@multishell nanowires and gathers some of the already mentioned advantages of the approach. It consists of the combination of different vacuum and plasma assisted deposition techniques, namely, physical vapor deposition of single-crystalline organic nanowires (step 1), magnetron or dc sputtering of metallic layers to form the shell acting as an inner electrical contact (step 2), and the plasma enhanced chemical vapor deposition (PECVD) of the polycrystalline piezoelectric shell (step 3). Such a multi-step procedure has evolved during the last five years for the exploitation of small-molecules single-crystalline organic nanowires (ONWs)[51] as vacuum-processable 1D and 3D soft-templates.[52-57] Among the advantages of this soft-template procedure, it is worth stressing the high compatibility with a variety of substrates since the formation of the ONWs is not dependent on catalytic processes but produced by a crystalline growth led by the self-assembly of pi-stacked molecules by Van der Waals forces.[52,58] To promote the crystalline formation under mild temperature and pressure conditions, the substrates shall present a certain roughness, as the surface of a metal, metal oxide thin films, nanoparticles, or organic and polymeric layers. Figure 1 shows examples of organic nanowires formed on different substrates, including metal oxide layers as ZnO (Fig. 1a-b), silver and gold nanoparticles on flat or metal oxide substrates (Fig. 1c-d), and O2 plasma pretreated graphene (Fig. 1e-f). The compatibility also extends to the supports, from flat Si wafers and polymers (as PDMS, PET)[49,52] to complex substrate morphologies and chemical compositions as cellulose, metallic meshes, and interdigitated electrodes (Fig. 1g-i).[57] Factors as substrate temperature and roughness, surface pretreatments (in the case of metal oxides and organic supports), deposition rate, time, pressure, and chemical structure of the small-molecules determine the density of ONWs, their length, thickness, and shape. Thus, densities in the range between 0.2 NW / $\mu m^2$ up to 30 NWs / $\mu m^2$ and lengths between less than 0.5 µm up to 20 µm can be achieved by tuning the experimental parameters and the molecule used as a building block.[51-58] Besides, it must be stressed that the ONWs present interesting properties as their molecularly flat surfaces, high mechanical flexibility, and semiconducting character.[59-62] These characteristics make them not only desirable as sacrificial soft-templates but also as an active part in the mechanical energy harvesters.

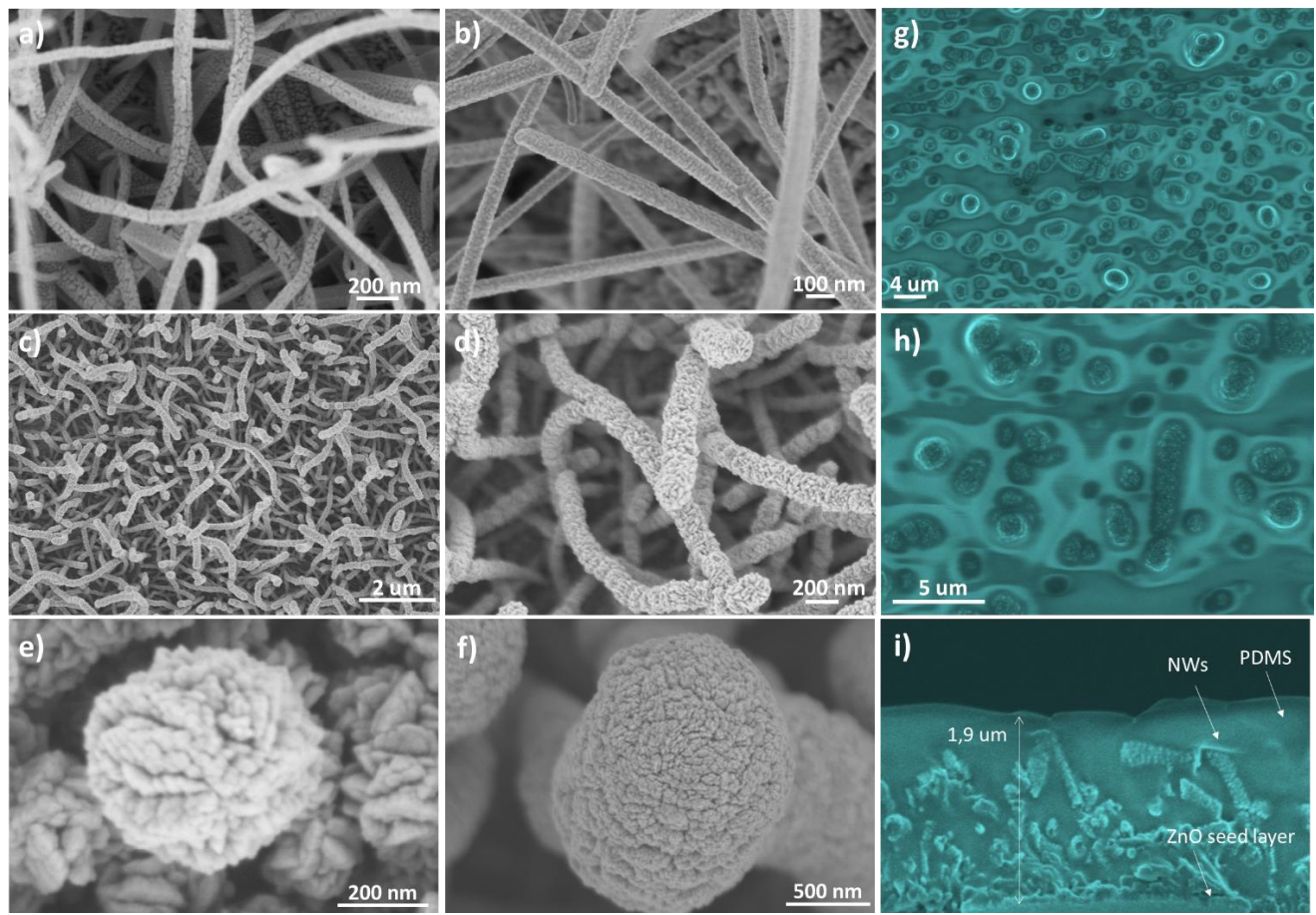

**Figure 2. Morphology and nanoscale microstructure of the core@shell NWs and embedded systems.** Characteristic SEM images of steps 2- 5) for different coalescence of Ag (a-b) and thickness of the ZnO shell (c-f). Panels on the right show the embedding of the $H_2Pc@Ag@ZnO$ NWs on PDMS in top view (g-h) and cross-section (i).

The second step in Schematic 1) addresses the formation of a metallic shell surrounding the organic core. We demonstrated in reference 49 that the introduction of this intermediated layer enhanced the current output of the PENGs by simply improving the conductivity of the core@shell nanowires and slightly increasing the rigidity of the system at the nanoscale in comparison with the ONW@ZnO nanowires. This latter characteristic allowed a more efficient mechanical deformation of the piezoelectric layer. The SEM micrographs in Figure 2 a-b) show the formation of the silver shell on the ONWs by DC sputtering for two different thicknesses. Panel a) demonstrate the formation of a non-coalescence silver layer for the equivalent thickness of silver below 50 nm, meanwhile in panel b) a completely connected silver shell is formed for an equivalent silver deposition above 100 nm (please note that the equivalent thickness, i.e. the thickness corresponding to a thin film layer deposited during the same experiment but without the 1D soft-template, is about 80 and 150 nm, respectively). It is interesting to address that even for low thickness shells, the percolation is quite high because of a fair wetting of the silver on the surface of the organic nanowires. It is also worth stressing that dewetting of the silver layer is not expected as the next step is carried out at room temperature. Besides, the DC sputtering method provides the growth with higher conformality than when applying thermal evaporation,[63] allowing the silver clusters to reach all the length of the 1D template, even the interface with the substrate. In the following, for the assembly of the piezoelectric nanoarchitectures, we will work with the percolated silver shells in panel b).

Step 3) is the formation of a piezoelectric shell by plasma enhanced chemical vapor deposition (PECVD). A critical advantage of the use of PECVD for the fabrication of metal oxide shells is the characteristic low temperature of processing and high conformality.[47-49,52-53] Figure 2 c-f) shows the formation of the ZnO shell on the 1D templates. The shell depicts a nanocolumnar morphology growing radially from the inner core and revealing a triangular shape at the surface of the NW. This morphology is characteristic of polycrystalline and porous ZnO layers as detailed elsewhere.[52,53] The medium thickness of the nanowires can be controlled by simply tuning the deposition time and the growth rate.

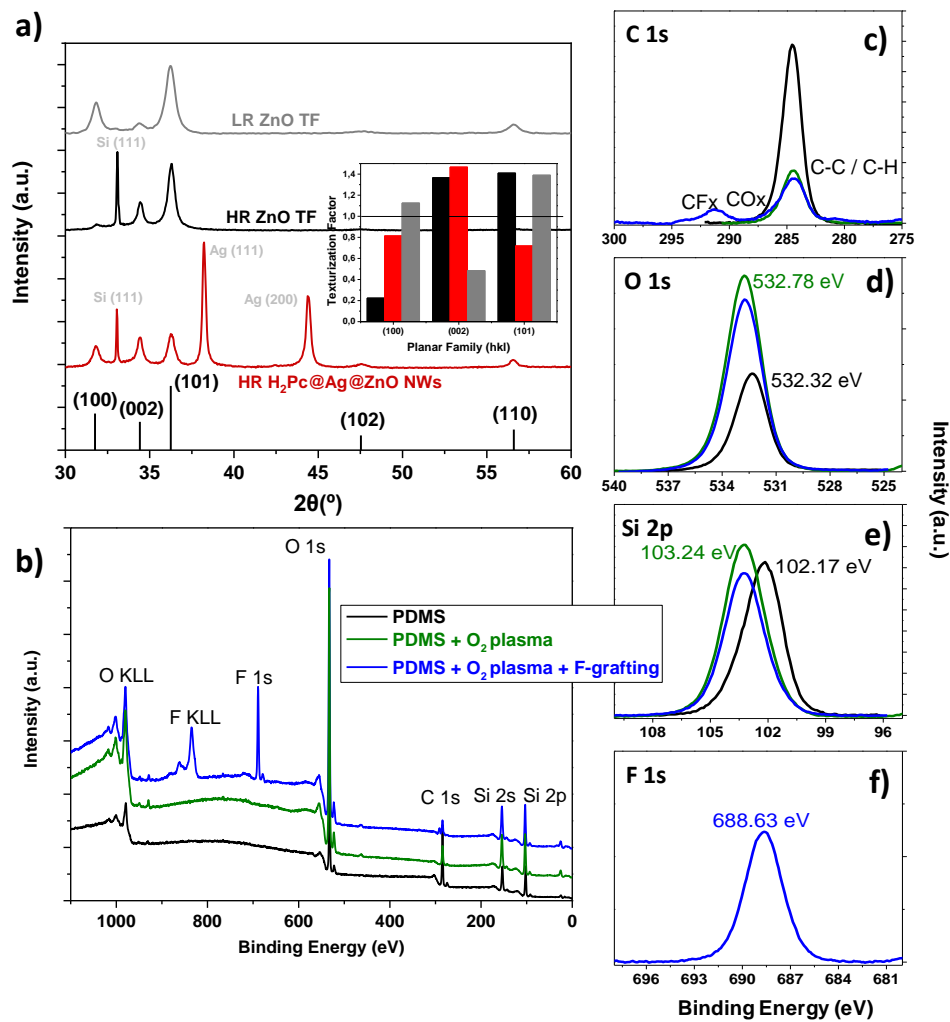

**Figure 3. Structural and chemical analyses of the devices.** (a) XRD of ZnO thin films grown at low (LR) and high Rates (HR) and core@shell nanostructures (H2Pc@Ag@ZnO) with the oxide layer grown at HR; XPS (general b) and zone peaks (c-f)) analyses analyses.

The ZnO shells of the NWs shown in panels c - e) were grown at a growth rate of below 5 nm/min, thereafter labeled as LR (low rate) conditions. As an example, panel f) presents a detailed view of nanowires grown at a higher growth rate of ca. 10 nm/min (HR), this time with an equivalent thickness higher than 900 nm. It is worth mentioning that the fabrication method also allows tuning the aspect ratio of the nanowires, as both, thickness and length can be varied separately. Such a feature will be exploited in the next section to produce hybrid nanogenerators with different responses.

When dealing with ZnO, the self-oriented growth along the c-axis of the wurtzite crystalline phase facilitates the exploitation of the single-crystalline nanowires as piezoactive systems since the growth direction of the wires corresponds to the anisotropic direction in the crystalline structure.[18-20] Under standard growth conditions by PECVD, the ZnO forms a polycrystalline layer as shown in the XRD patterns in Figure 3 a) and elsewhere.[47-49, 64] In polycrystalline thin films, what determines the preferential formation of crystalline planes is the degree of texturization.[64,65] Thus, the Texturization factors, T(hkl), extracted from the X-Ray Diffraction patterns give a quantitative appreciation of the preferential formation of crystalline planes parallel to the substrates. The structure and texture of metal oxide thin films can be controlled through several experimental factors, being the substrate temperature and the plasma gas composition the most usually tuned.[47] However, for the particular application of these polycrystalline layers within nanogenerators assembled in polymeric substrates, we propose herein to explore the enhanced texturization obtained by growing at high growth rates. This approach improves the piezoresponse of the nanowires at the same time increases the competitiveness of the PECVD method in front of other alternative routes as high yield synthetic protocols are always welcome looking to the eventual scalability of the process. Several articles have previously settled the enhanced texturization of polycrystalline metal oxides layer as wurtzite and anatase grown at high growth rates. Indeed, we demonstrated in reference 65 the growth of highly texturized PECVD layers of anatase in the form of faceted columns following the Kolmogorov mechanism. In the same way, the transport mechanism of two different texturized ZnO layers was discussed in comparison with ZnO nanoparticles for their implementation in dye sensitized solar cells,[64] depicting different mechanisms but parallel recombination kinetics and electron collection efficiencies. Figure 3 a) gathers the XRD patterns and textural factors (inset) for thin film layers grown under low (LR) and high growth rates (HR) and for the core@multishell nanowires (H2Pc@Ag@ZnO) grown at high growth rates. Following the calculation gather at the Experimental Section, values higher than 1 for T(hkl) indicates a preferential texturization for that plane. The texturization factors for LR conditions depict a preferential orientation on planes (100) and (101), meanwhile, the samples grown under HR present a preferential texturization for (002) planes, with a slightly higher value for the NW sample. The crystal sizes were estimated using the Scherrer formula in nanometer size as gathered in Table S1 at the SI for the three samples and all the planes presented in the XRD patterns. The crystal sizes are always below 40 nm, with those corresponding to the nanowires slightly smaller than the thin film ones.

Step 4) consists of the embedding of the ZnO nanowires in a PDMS matrix by standard spin coating of the polymer in a toluene solution (see Experimental Section). The as-prepared NWs sample is superhydrophilic as a result of the hydroxylation and reactivity of the ZnO surface grown under oxygen plasma[66-68] combined with the aspect ratio of the NWs which increases the surface roughness. Therefore, the PDMS solution easily spreads on the surface, following a Wenzel wetting behavior and filling the voids between the NWs as shown on the SEM micrographs in Fig. 2 g-i). The thickness of the PDMS matrix can be controlled to completely cover the NWs' length or to leave them partially exposed as in panels (g-h). Although, even in this latter case, a thin layer of PDMS is formed at the surface of the NWs and entering into the open pores and smoothing their surface (see Figure S1 at the SI section). In the last steps, 5) and 6), the surface of the PDMS is pretreated by oxygen plasma to provide the chemical binders for the fluorination with the PFOTES molecules (hereinafter F-grafting).[57,69] The effective surface perfluorination of the PDMS was followed by X-ray Photoelectron Spectroscopy (XPS) by analyzing the chemical composition of the PDMS surfaces at the different stages of the conditioning and grafting (see Figure 3 b – f and Table 1).

**Table 1.** Elemental atomic concentration obtained by XPS analysis and WCA corresponding to the NWs sample completely covered by PDMS.

|  | C (at %) | O (at %) | Si (at %) | F (at %) | O/Si | F/Si | WCA |
| --- | --- | --- | --- | --- | --- | --- | --- |
| PDMS | 43 | 28 | 29 | - | 0.97 | - | 90º |
| PDMS + O$_2$ plasma | 11 | 57 | 32 | - | 1.78 |  | <10º |
| PDMS + O$_2$ plasma + F-grafting | 12 | 49 | 26 | 13 | 1.88 | 0.50 | 106º |

The general spectra in panel b) show the overall composition of the surfaces, with the presence of C, O, and Si in all the cases. It also confirms the presence of F on the grafted sample. The peak zone analysis (Figure 3 c-f) allows for the comparison of atomic percentages and brings information about the chemical state. Thus, the oxidizing plasma treatment doubles the surface oxygen content as Table 1 indicates through the significant carbon content decrease and the O/Si ratio from 0.97 to 1.78. This chemical modification is reflected in a shift to higher binding energies as it is observed in panel d), which indicates the predominance of Si–OH over Si–O–Si functional groups. It is supported by the panel e) where the Si2$^+$ (–[Si(CH3)2–O–]–, the main contribution of the PDMS surface at 102.2eV, is also shifted to higher binding energies around 103.2eV corresponding to Si3$^+$ species in [Si(CH3)(OH)–O–]– bond configuration.[37] It is also worth noting the negligible intensity of the C1s spectrum (panel c) at the binding energies of CO$_x$ groups, which agrees with the preferential formation of new Si-O bonds after the Si-CH3 bond scission. This finding along with the > 80 % increase of the O/Si ratio (O/Si$_{plasma}$ = 1.83 O/Si$_{PDMS}$) points to the almost complete oxidation/hydroxylation of the surface Si atoms as a consequence of the plasma pretreatment. It is worth mentioning here that such complete oxidation occurs at RT after treatment of only 15 minutes under a plasma power of 600 W, i.e. less than the usual consumption of the typical home appliance. It is worth stressing herein that the use of oxygen plasma in such pretreatment is an environmentally friendly alternative to the standard procedures requiring the surface activation with hydrogen peroxide or acid followed by the exposure to PFOTES by reflux in toluene.[70-71] Besides, contrary to previous reports in the literature on the plasma treatment of triboelectric polymers,[34-36,38] the surface roughness of the PDMS is barely modified as shown in Figure S1 at the SI. The implications of such a feature are twofold, on one hand, the mechanical and optical properties of the PDMS will be not affected by the plasma treatment; on the other hand, the enhancement of the charge surface generation can be univocally linked to the F-grafting and separated from the increase of the contacting surface area. Finally, the results in Fig. 3 (c, f) demonstrate the perfluorinated grafting is well performed after the plasma treatment as addressed by the fluorine content and F/Si ratio of 0.5 in Table and the increment in C consequently with the chemical composition of the PFOTES molecules. This occurs through the reaction between the –CH3 at the siloxane head of the PFOTES molecule with –OH groups on the plasma activated surface.[69] Hence, the F1s band at 688.6 eV is a signal of the bond between fluorine and carbon atoms as well as the second main band appearing in the C1s spectrum at 290.7 eV corresponding to $CF_x$ species respectively.[72]

As indicated above, the WCA of the triboelectric layer determines the lifespan of the charges generated by triboelectrification. Thus, for higher WCA is expected an improved stability of the triboelectricity as the surface intrinsically avoids the detrimental attenuation effect given to water molecules and hydroxyl groups.[21,29,39] Moreover, the fluorination of the surface is expected to enhance the performance of the negative triboelectric layers as the fluorine shows the highest affinity for electrons.[29,39] The WCA value for the PDMS embedded NWs is ∼ 90º. The plasma oxygen treatment reduces this CA drastically consequently (< 10º) to the formation of a SiO2 layer and –OH groups covering the PDMS. Finally, WCAs increases (ca. 106 º) are found after the F-grafting with PFOTES in good agreement with previous recent articles on the development of hydrophobic PDMS triboelectric layers. Please note, that for this comparison we characterized the fully embedded nanowires samples, with characteristic SEM micrographs in Figure 2 i). When dealing with a thinner layer of PDMS, as the example in Figure S2, the WCA of the fluorinated surfaces dramatically increases up to superhydrophobic values as a consequence of the combination of low surface tension given by the F-molecules and the high roughness produced by the micrometer length of the nanowires. It is also worthy to mention that this approach might provide a route to produce self-cleanable triboelectric layers[57] and to allow the functioning of the hybrid piezo/triboelectric nanogenerator even at low temperatures, as the F-grafting has been demonstrated to delay the freezing of droplets and water condensation under sub-zero conditions.[69]

*Power generation and characterization of the piezoelectric and triboelectric hybrid nanogenerators.*

Figure 4 gathers the voltage (load of 10 MΩ) and short circuit current obtained for a PTHNG built following the architecture at the right side of Schematic 1. The NWs were formed on a thin layer of ZnO deposited on commercially available ITO/PET and then coated with PDMS. When an external force is applied to the device, like the fingertip force by tapping, the top electrode moves in the perpendicular direction to the device surface and the triboelectric layer and PET interface are brought into contact at the same time that the whole system is bent. The piezoelectric layer is therefore activated by direct pressing under the fingertip but also by the bending of the PET substrate. In this way, the piezoelectric and triboelectric mechanisms work together in a single pressing and releasing cycle. Fig. 4 a and b) shows the comparison for the hybrid nanogenerators at different levels of optimization. Thus, the lowest performance for both $V_{10M\Omega}$ and $I_{sc}$ corresponds to the PDMS device, fabricated by direct spin-coating deposition on the ZnO seed layer (without NWs) (black lines). The following one in the rank is the $H_2Pc@ZnO$ NW device formed under a low growth rate of the ZnO shell (grey lines) which shows an appreciable improvement in both voltage and generated current. Such an enhancement is surpassed once the nanowires are modified by including the inner conducting silver shell and increasing the texturization along the (002) direction of the ZnO shell (cf. Fig. 3 a), i.e. $H_2Pc@Ag@ZnO$ NWs (yellow line). In the previous reference [49], we demonstrated the enhanced response of the PENGs fabricated with an inner gold shell. The system multiplied by 30 the output current generated by the $H_2PC@ZnO$ NW PENGs because of the improvement in the conductivity and by slightly increasing the rigidity of the system at the nanoscale in comparison with the ONW@ZnO nanowires. This latter characteristic allowed a more efficient mechanical deformation of the piezoelectric layer. However, in the best of these PENGs, the highest value obtained for Isc was ca. 60 nA with output voltage in the range of 100 mV, much smaller than the obtained for the PTHNGs in Fig. 4 with current as high as 0.2 µA and voltages reaching 20 V. The final step by chemical modification of the triboelectric layer with fluorine grafting improves the previous values with a major impact in the output voltage which reaches almost 80 V by fingertip actuation with a frequency of approximately 3 bps (beats per second). The current output is also enhanced with peaks as high as 0.27 µA. This output is enough to power commercial light emitting diodes LEDs as demonstrated in panels c) and d) directly connecting the PTHENG to the LED array (Fig. 4 c) or by charging a 500 nF capacitor through a diode bridge (Fig. 4 d).

Figure 5 shows the proposed dynamics of the piezo and triboelectric hybrid device during one cycle of operation. Because of the repeated contacts between the ITO/PET electrode and the F-grafted layer, a relatively high density of triboelectric charges accumulates on the surface of the triboelectric layer. When the vertical force is applied on the top electrode (Fig. 5a) electric current $i_L$ flow through the load from the bottom electrode to the top. As the gap is closed and a contact forms between the top electrode and the dielectric layer, the applied vertical force reaches the maximum value of $F=F_{max}$ (Fig. 5c). The PDMS Young's module depends on the curing agent ratio but it is expected to be always smaller than the corresponding to the Ag@ZnO shells, even though our NWs still contain the organic core. Thus, when the force is applied to the device, the deformation occurs firstly in the PDMS and then is delivered homogeneously to the ZnO, acting as a diffuser layer that applies the stress isotropically to the NWs.[73,74]

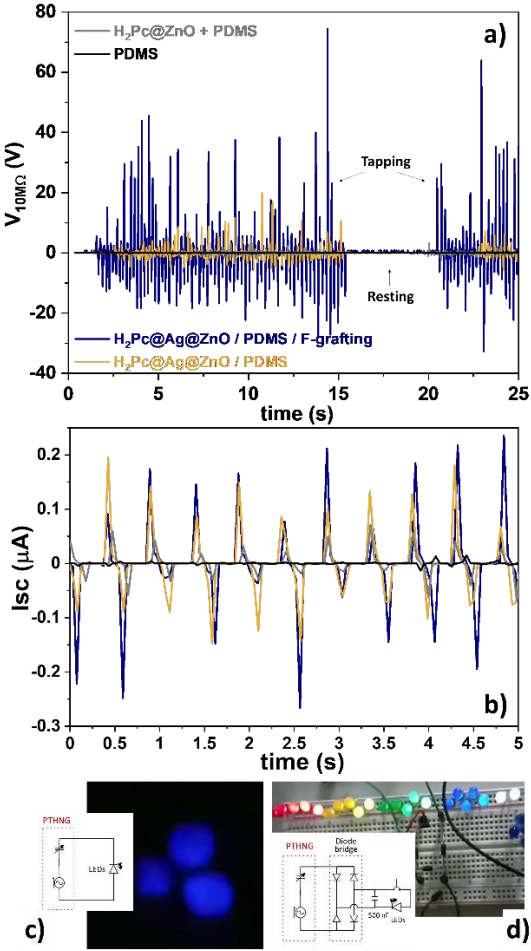

**Figure 4. PTHNGs activation by tapping for medium aspect ratio NWs.** Increment in voltage (load of 10 MΩ) (a) and shortcircuit current (b) following the optimization of the architecture for four different synthetic steps as labeled. c-d) Proof of concept as a power source for LEDs for direct lighting (c) and by charging a microcapacitor (d).

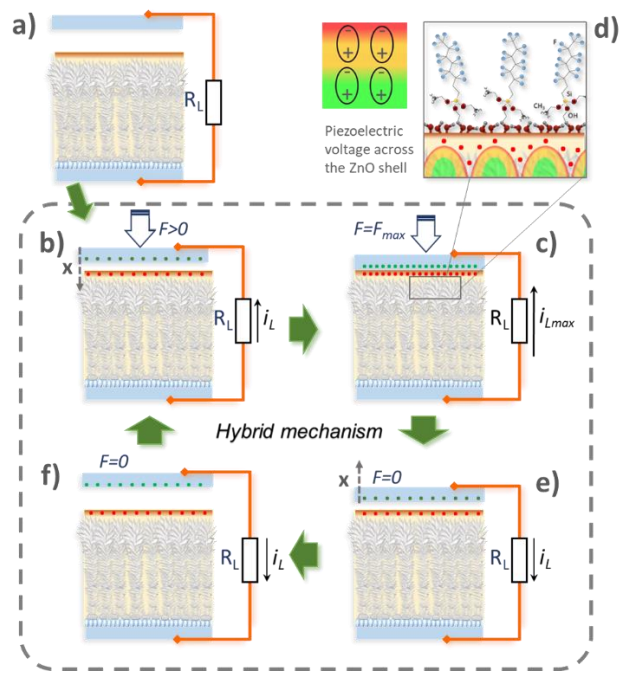

**Figure 5. The mechanism proposed for the PTHNGs.** a) the hybrid device in the rest position with no vertical force applied, b) device in the actuated state with a load current of $i_L$, c) piezoelectric polarization is generated with maximum applied force resulting in a higher load current; d) generated negative charge on nanowires and negatively charged fluorine molecules; e) and f) the applied force is removed and the device goes back to the rest state and load current flows in the opposite direction.

Thus, because of the presence of this force in the contact state, the ZnO nanowires are compressed from all sides and induce a negative piezoelectric polarization on the outer shell (Fig. 5d). At the same time, the fluorinated grafting attracts free electrons during repetitive contact with the top electrode. Eventually, both bound and free negative charges due to piezoelectric polarization and triboelectric contacts, respectively, add up together and initiate a higher charge induction on the electrodes. The combined effect of both contributions increases the external flow of electrons necessary to compensate for the mechanically induced charge distribution. Thus, under this condition of maximum deformation and contact between electrodes, the load current reaches a maximum value of $i_L = i_{Lmax}$. As the contact breaks, the effect of piezoelectric polarization vanishes, and the gap between the electrodes increases. This reverses the flow of the load current, now from the top to the bottom electrode, to keep the electrostatic neutral state (Fig. 5e and 5f).

We have gone a step forward and carry out a thorough characterization of the power output of the PTHNGs by using a magnetic shaker and under a precisely controlled frequency (Figure 6). For such analysis we took advantage of the flexibility of the synthetic protocol to produce nanowires with different lengths and thickness so we constructed PTHNGs with piezoelectric nanowires characterized by three different aspect ratios (AR): low (low AR ~ 1.7, total thickness in the range of 900-1200 nm, length between 1.3 and 2.3 µm, equivalent ZnO thickness of 900 nm), medium (medium AR ~ 3, medium total thickness about 450 nm, length in the range of 0.7 to 1.7 µm, equivalent ZnO thickness 200 nm) and high (high AR ~ 8.3, medium total thickness about 400 nm and lengths reaching 3.5 and 4 µm, equivalent ZnO thickness 400 nm). Please note that the label in the panels a) to c) indicates the AR and also the equivalent thickness of ZnO, i.e. the thickness of the ZnO thin film deposited under the same experimental conditions but on a flat substrate. The shaker is adjusted to impose different vertical forces onto the top electrode with the same magnitude around 0.75 N, but with various frequencies. Fig. 6 a-c) shows the curves for dissipated power in a variable resistive load at different activation frequencies (4, 12, and 20 Hz). At a first sight, the three devices show the same trend regarding the response to the frequency, as for all of them, the power density increases with frequency in the selected range. On the other hand, we find a strong dependency on the aspect ratio, with a maximum power density of near 6 µA/cm$^2$ obtained for the medium aspect ratio NWs (Fig. 6b) with the smallest response corresponding to high AR nanowires (Fig. 6 c). Besides, in the case of low and medium AR, power density curves show a shift to the left at higher frequencies as it is also typically observed and theoretically analyzed in single piezoelectric and triboelectric nanogenerators.[74-75] Meanwhile, for the case of long, high AR NWs the optimum load for actuation frequency of 4 Hz is clearly smaller compared to higher frequencies which make this device highly compatible with very low-frequency applications such as energy generation from human body movements. This is likely connected to the fact that NWs with a higher aspect ratio have more room for flexion and bending in response to the applied force. So concerning the proposed mechanism in Fig. 5, at low frequency, for the high AR nanowires, the activation of the device does not only produces the homogenous compression of the ZnO shell through the PDMS but also a non-negligible strain because of the bending of the NWs. This is likely the reason why the best performace corresponds to the medium AR NWs, where the thickness of the ZnO shell and the length of the nanowires allow for both pressing and strain of the piezoelectric counterpart.

Although the direct comparison with output powers from another mode I PTHNGs is hindered by the variety of materials, architectures, and operating conditions. Our results in Fig. 4 and 6 (a-c) are in good agreement with the trend in literature including a huge enhancement of voltage output when comparing simple PENG and PTHNG and the increment in current output for piezoelectric, ferroelectric and multiferroic nanomaterials embedded in the triboelectric layer.[21-28, 76-77] In our case, we can also link the AR of the NWs with the output power and response to different frequencies, showing that for thin equivalent thicknesses of the ZnO and therefore nanoscale shell thicknesses, the response of the hybrid system is already defined by the presence of the piezoelectric fillers.

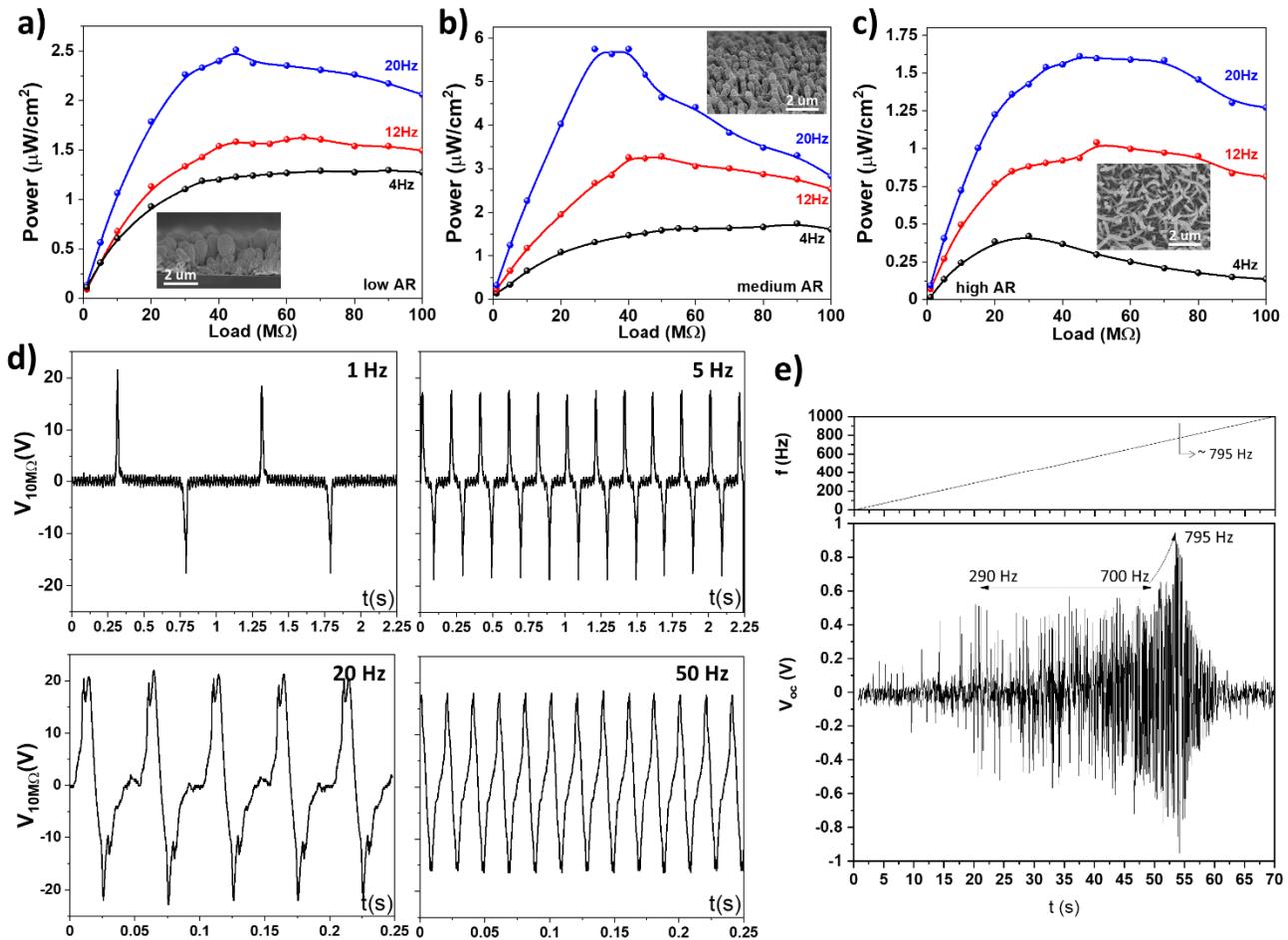

**Figure 6. The power density of the PTHNGs and response to different frequencies.** a) - c) Power density versus load curve for devices fabricated with NWs characterized by three different aspect ratios as labeled. d) $V_{10M\Omega}$-t curves corresponding to the medium AR NWs in Fig. 4 and panel b) activated by the magnetic shaker at four different frequencies. e) Response of the medium AR NWs PTHNG to acoustic vibration actuation for increasing frequencies (without pressing the top contact electrode).

Interestingly, the nanogenerators positively respond to the different imposed frequencies of the shaker, even though the selected range is ample. The literature[6-15] gathers two main approaches to produce low-frequency wide bandwidth energy harvesters, relaying on multimodal or nonlinear wide-band frequency vibration methods. The multimodal energy harvester increases the harvesting bandwidth by adding nanogenerators with different resonance frequencies, usually leading to large size devices. Typical nonlinear energy harvesters work in cantilever configurations including one or more stoppers.[6-15,75] With the results in panels d and e) in Fig. 6 for the medium AR NW devices, we aim to demonstrate that even for a very simple configuration in vertical contact mode and only one type of piezoelectric fillers, our PTHNG cover a wide range of operating frequencies for direct actuation under the shaker (panel d) and even by just leaving the system on the surface of a speaker (panel e). Thus, the V/t obtained for the load corresponding to the connections to the oscilloscope, i.e., 10MΩ, the output reaches peak to peak values ca. 20 V for the range between 1 to 50 Hz, which can be considered as ultrawide bandwidth when comparing with other piezo and triboelectric hybrid solutions in the literature.[6-15] Outstandingly, the response to the direct acoustic vibration is homogeneous in a larger range between 290 and 800 Hz, reaching a maximum of output voltage about 795 Hz. Please consider that for this characterization, the PTHNG was gently emplaced on top of the surface of a speaker without any imposed vertical force which is the reason why the output voltages are much smaller than those in panel d). A likely explanation is that in the case of sound activation, pressure waves are applied to the bottom ITO/PET substrate, which are transfer to the ZnO shell through the PDMS deformation producing the excitation of the piezoelectric response. On the other hand, the acoustic (pressure) waves would also interact with the top ITO/PET electrode in a rather similar way. This is, both ITO/PET layers would experience a similar deformation. Consequently, the ITO/PET and the F-grafted internal surfaces of the nanogenerator are not brought into close contact by this gentle stimuli, and we can expect that a non highly effective charge transfer is induced between these surfaces

Finally, at this point, it is worth stressing that all the V/t, I/t, and output power densities characterizations gathered in Figures 4 and 6 were carried out under relative humidity higher than 50%, as the room conditions for our laboratory. Given the COVID pandemic situation, devices were stored for several months under such conditions, and experiments were repeated to ensure reproducibility with non-detriment on the electrical properties neither on the water contact angles of the F-grafted triboelectric layers. Thus, these early results altogether with previous reports in the literature support the advantages of the functionalization of PDMS layers with F-molecules not only for the enhancement of the surface charge capacity but also to produce robust and long-lasting devices.

**CONCLUSIONS**

Herein we have shown the step-by-step fabrication of a piezo and triboelectric hybrid nanogenerator demonstrating the advanced application of plasma-assisted technologies not only for the pretreatment of elastomeric PDMS but also for the growth of highly texturized piezoactive ZnO layers and conducting silver shells. The application of plasma-assisted technologies to the treatment and conditioning of polymers has been exploited for decades on a multitude of topics ranging from food packaging to biomaterials and agriculture, from microfluidics to microelectronics with two paramount advantages: their high scalability and environmentally friendly character.[31,76-78] Thus, these techniques have granted compatibility with extended fabrication protocols as those presented in microelectronics and roll-to-roll technologies.

We have demonstrated the compatibility of the synthetic method with a wide variety of substrates and supports, ranging from ITO/PET to metal meshes, cellulose or interdigitated electrodes. Among the several steps of optimization of the hybrid mode I nanogenerators, the best performance (power density values up to 6 µW/cm$^2$) has corresponded to the device with medium aspect ratio nanowires and a core@multishell architecture including highly texturized ZnO and conductive Ag shells embedded in the F-grafted PDMS. XPS and SEM results have demonstrated the almost complete conversion of the PDMS surface after the oxygen plasma treatment to oxidized Si allowing for an enhanced anchoring of the PFOTES molecules while keeping the smooth roughness characteristic of the elastomeric polymer. The PDMS functionalization by F-grafting plays a twofold role, improving the charge generation at the surface of the polymer and increasing its hydrophobicity by almost 20º. This latter feature allows for highly stable, reproducible, and durable devices.

A hybrid mechanism taking into account the charges created by the intermittent contact of the PDMS triboelectric layer and the ITO/PET electrode and the piezoelectric potential arise by the isotropic pressing of the ZnO shells and wires through the PDMS has been proposed. This is hybrid excitation mechanism is supported by i) the efficient piezoelectric response reported for core@multishell ZnO nanowires (see ref. [49]), and ii) the enhanced triboelectric effect demonstrated herein for the F-grafting in the complete devices. The response of the PTHNGs has been positively evaluated under wide bandwidth frequency vibrations with power densities values slightly increasing for low-frequency ranges (from 1 Hz to 50 Hz). At higher frequencies, the output voltage is smaller given the experimental set-up applied for the characterization. However, it is shown how the system successfully responds to vibrations with frequencies up to ca. 800 Hz. The hybrid piezo and triboelectric mechanism activated by direct manual tapping generates power enough to feed an array of LEDs.

These results open a new path for the fabrication of hybrid piezo and triboelectric nanogenerators with elements compatible with the different modes for the hybrid systems and easily extendable to other piezoelectric and triboelectric materials. As a final remark, it is worth stressing that the application of tailored microstructured (and porous) piezoelectric shells within the PTHNG architecture provides a straightforward approach also to the development of self-powered nanosensors to agents as UV light, pollutants, VOCs, and hazardous gases.

**EXPERIMENTAL METHODS**

The main steps required for the formation of a high density of the core@shell nanowires are gathered in Schematic 1, namely, 1) formation of the seed layer (polycrystalline ZnO thin film or Ag nanoparticles); 2) growth of the ONWs with different lengths and density; 3) fabrication of the conformal Ag and ZnO shells by DC sputtering and PECVD respectively. The following steps include embedding in PDMS and its plasma-assisted functionalization.

*Formation of the ZnO nucleation seed layer and piezoelectric shells.* ZnO thin films and shells were fabricated by Plasma Enhanced Chemical Vapor Deposition (PECVD) at room temperature following an already established protocol. Diethylzinc (($CH_2CH_3)_2Zn$) from Sigma-Aldrich was used as delivered as Zn precursor. The oxygen and Zn precursor were delivered in the plasma reactor by using a mass flow controller. The base pressure of the chamber was lower than $10^{-4}$ mbar, the total pressure during the deposition was around $10^{-2}$ mbar. The plasma was generated in a 2.45 GHz microwave Electron-Cyclotron Resonance (ECR) SLAN-II operating at 800 W and the deposition was carried out in the downstream region of the reactor.

ZnO layer was deposited on the samples through a mask, leaving a free area of the substrate to emplace the Cu wires and electrodes (3M Cu adhesive tape nominal resistivity 0.005 Ω/sq). A thickness of 150 nm was settled for the seed layers deposited at 15 nm/min. Shells were deposited under different conditions ranging from 15 to 5 nm/min (high/low growth rate correspondently) and with thicknesses between 200 to 900 nm.

*Formation of the Ag conductive shells.* The Ag deposition was carried out by DC sputtering on an argon atmosphere. The base pressure of the chamber was lower than $10^{-2}$ mbar, the total pressure during the deposition was $1.6\cdot10^{-1}$ mbar of Ar and the voltage for the plasma generation was 600 V. The deposition time was varied to form a thin Ag coalesced layer.

*Growth of organic nanowires.* Organic nanowires (NWs) were fabricated by vacuum deposition of H2-Phthalocyanine (Sigma-Aldrich) using a low-temperature sublimation cell. The base pressure of the chamber was lower than $10^{-5}$ mbar. The total pressure during deposition was around 0.001 mbar of Ar, and the substrate temperature was 240 ºC. The sample-to-evaporator distance was 6.5 cm. To produce short nanowires the deposition rate in the QCM was first settled at 0.4 Å/s for 10 minutes and then increased up to 3 Å/s for a total equivalent thickness in the QCM of 1 kÅ. The long nanowires were deposited at growth rates higher than 1 Å/s for the entire deposition time until reaching an equivalent thickness of 1.5 kÅ.

*Implementation of the piezoelectric NWs in the TENG.* The as-grown NWs were embedded in polydimethylsiloxane (PDMS). First, the monomer was mixing with a curing agent (Sylgard 184 Sigma Aldrich) in a 2:1 ratio, and then the mixture dissolved on toluene, in a ratio 1:2 for thicker coatings (vertical separation-contact devices) or 1:3 for thin ones. A sufficient solution to cover completely the NWs was spin-coated and it was heated at 120 ºC for 3 h to cure the polymer. Spin coating was performed in a WS-400-6NPP-LITE coater from Laurell at 7500 rpm during 45 s.

*Fluorocarbon Grafting of PDMS.* The chemical derivatization by fluorocarbon grafting was carried out by exposition of the sample at 80 ºC to 1H,1H,2H-Perfluorooctyltrietoxisylane (PFOTES) 98% vapor on a static low vacuum (obtained with a rotatory pump) at least for 2 hours. To improve the effectiveness of this treatment, the samples had been previously softly etched 15 minutes in O2 plasma to generate a SiOH surface functionalization, as in Lopez-Santos et al.[57,69] This plasma pretreatment was carried out in the same ERC reactor used for the PECVD deposition but, in this case, the microwave power was 600 W.

*Supports and devices configurations.* The configuration selected for the TENGs was a vertical contact-separation mode, using commercially available ITO/PET substrate purchased from Sigma Aldrich (60 Ω/sq) and covered with the core@multishell architecture. The second contacting surface of the device was an ITO/PET sheet of the same characteristics as the previous one but used as delivered. A piece of copper tape and a thin wire was attached to the end of each electrode where the ITO layer was exposed. Both electrodes were assembled using Kapton tape so that several turns of the tape produced a gap of a few millimeters along the ends of the electrode and also isolated the copper part.

*Characterization of the obtained structures and devices testing.* SEM micrographs were acquired in a Hitachi S4800 and in a Teneo from FEI working at 2 kV at working distances in the range of 2-4 mm. The crystalline structure was analyzed by X-Ray Diffraction (XRD) spectrometer in a Panalytical X'PERT PRO model operating in the θ - 2θ configuration and using the Cu Kα (1.5418 Å) radiation as an excitation source. The texture coefficients T(hkl) were calculated applying the following equation:

$$T(hkl) = \frac{I(hkl)/I_0(hkl)}{\frac{1}{n}\sum I(hkl) / \sum I_0(hkl)}$$

Where I(hkl) and I$_0$(hkl) are the peak intensities associated with the (hkl) family plane obtained in Bragg-Brentano (Ө-2Ө) configuration for the samples and the randomly oriented reference pattern respectively (in this particular case the JCPDS card Nº 36–1451, for wurtzite) and n is the number of possible reflections. If the texture coefficient is greater than 1, the studied samples have a preferential planar growth in that direction in comparison with the reference. Surface chemical composition was carried out by X-ray Photoelectron Spectroscopy (XPS) with a PHOIBOS 100-DLD spectrometer working in the constant pass energy mode fixed at a value of 20 e.V Spectra were recorded with the AlKα line and the binding energy (BE) scale was referred to the C1s peak at 284.5 eV for the C-C/C-H functional groups of the surface. Water contact angle measurements were performed in an OCA20 goniometer from Dataphysics with 2 µl droplets of bidistilled water. The values showed in Table 1 correspond to the mean values of at least 5 droplets.

The voltage measurements were performed for different loads, using a digital oscilloscope (Tektronix TDS1052B) in the case of the vertical separation-contact devices. Firstly, mechanical stimuli were produced by pressing manually the surfaces. Secondly by connecting the devices to a magnetic shaker (Smart Shaker K2007E01) attached to a force sensor (DYTRAN 6011A). This combination provides the acquisition of curves with a more reproducible character of the peaks (both in amplitude and frequency), and also for properly characterizing the output as a function of the frequency, and for increasing loads. The output power is estimated by averaging the power between three cycles of mechanical actuation. The response of the medium AR nanowires device to acoustic stimuli was analyzed with the help of a pair of speakers (Trust 19830-02). Each device was placed in the center of each speaker and fixed with adhesive tape. Different waves were generated and the response of the PTHENGs was measured. The current measurements were performed in short-circuit, connecting the electrodes of the PTHENG to a Keithley 2635A.

## ACKNOWLEDGMENTS

We thank the AEI-MICINN (PID2019-110430GB-C21 and PID2019-109603RA-I0), the Consejería de Economía, Conocimiento, Empresas y Universidad de la Junta de Andalucía (PAIDI-2020 through projects US-1263142, ref. AT17-6079, P18-RT-3480), and the EU through cohesion fund and FEDER 2014–2020 programs for financial support. JS-V and CLS thank the University of Seville through the VI PPIT-US. JS-V also thanks to the Ramon y Cajal Spanish National programs. We thank the XPS service from ICMS and the CITIUS from the University of Seville for advanced characterization. The project leading to this article has received funding from the EU H2020 program under grant agreement 851929 (ERC Starting Grant 3DScavengers).

# SUPLEMENTARY INFORMATION

# Plasma engineering of microstructured piezo – triboelectric hybrid nano-generators for wide bandwidth vibration energy harvesting

Xabier García-Casas,[1] Ali Ghaffarinehad,[1,*] Francisco J. Aparicio,[1,*] Javier Castillo-Seoane,[1,2] Carmen López-Santos,[1,3] Juan P. Espinós,[1] José Cotrino,[1,2] Juan Ramón Sánchez-Valencia,[1,2] Ángel Barranco[1] and Ana Borrás.[1*]

1 Nanotechnology on Surfaces and Plasma Group (CSIC-US), Materials Science Institute of Seville (Consejo Superior de Investigaciones Científicas – Universidad de Sevilla) c/Américo Vespucio 49, 41092, Sevilla, Spain.

2 Departamento de Física Atómica, Molecular y Nuclear, Universidad de Sevilla, Avda. Reina Mercedes, E-41012, Seville, Spain.

3 Department of Applied Physics I, the University of Seville, Virgen de Africa 41011 Seville, Spain.

**Table S1.** Crystal sizes as estimated by the Scherrer equation from the XRD patterns in Figure 3.

| Plane | Peak Position (º) | FWHM | Crystallite size (nm) |
|---|---|---|---|
| ZnO Thin film Low Rate | | | |
| (100) | 31.84 | 0.318 | 37.9 |
| (002) | 34.46 | 0.340 | 34.6 |
| (101) | 36.31 | 0.478 | 22.1 |
| ZnO Thin film High Rate | | | |
| (100) | 31.81 | 0.396 | 27.9 |
| (002) | 34.37 | 0.356 | 32.4 |
| (101) | 36.25 | 0.475 | 22.3 |
| Nanowire with ZnO Shell High Rate | | | |
| (100) | 31.77 | 0.475 | 22.0 |
| (002) | 34.41 | 0.396 | 28.1 |
| (101) | 36.31 | 0.554 | 18.4 |

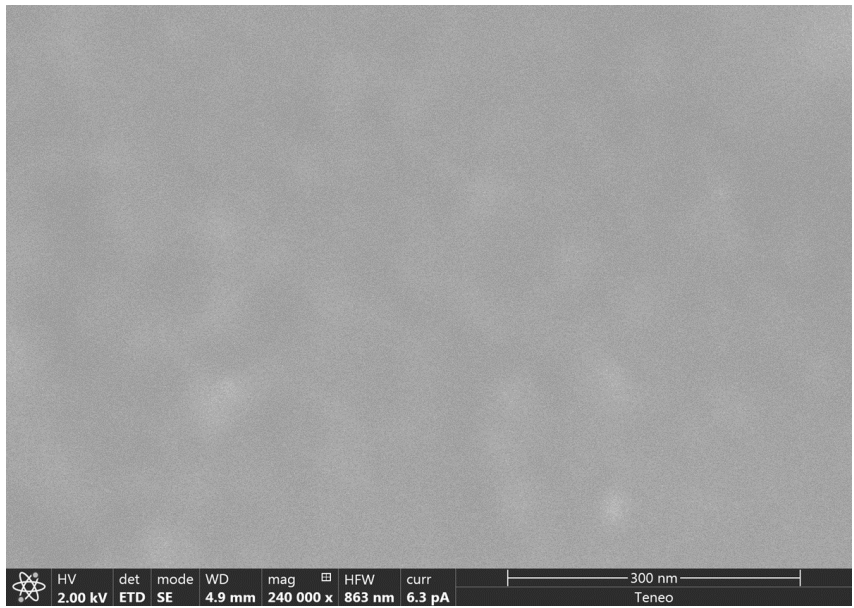

**Figure S1.** SEM micrograph of the PDMS surface after the oxygen plasma treatment carried out at 600 W in a MW-ECR reactor at $10^{-2}$ mbar of oxygen for a time of 15 minutes.

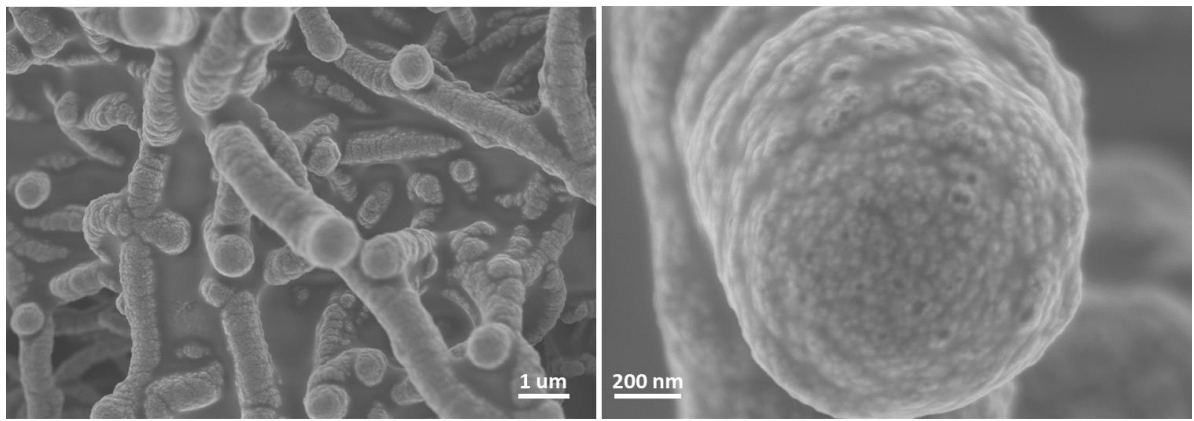

**Figure S2.** SEM images of core@multishell nanowires embedded in a thin layer of PDMS acting as an external coating of the nanoarchitectures. Please note that the PDMS completely fills the pores and empty spaces present in the ZnO shell.

33